\documentclass[preprint,showpacs,preprintnumbers,amsmath,amssymb]{revtex4}

\usepackage{graphicx}% Include figure files
\usepackage{dcolumn}% Align table columns on decimal point
\usepackage{bm}% bold math

\newcommand\GeV{\,\mbox{GeV}}

\begin{document}
\title{Nonsinglet parton distribution functions from the precise next-to-next-to-next-to leading order
QCD fit}
\author{Ali N. Khorramian $^{a,b}$}
\email{Khorramiana@theory.ipm.ac.ir}
\homepage{http://particles.ipm.ir/}
\author{H. Khanpour $^{a}$}
\email{hamzeh_khanpour@nit.ac.ir}
\author{S. Atashbar Tehrani $^{b}$}
\email{Atashbar@ipm.ir}
\affiliation{%
$^{(a)}$ Physics Department, Semnan University, Semnan, Iran}
\affiliation{%
$^{(b)}$ School of Particles and Accelerators, Institute for
Research in Fundamental Sciences (IPM), P.O.Box 19395-5531,
Tehran, Iran}
%This line break forced with \textbackslash\textbackslash
%}%
%\preprint{IPM/07-05-2009-SPA}
\date{\today}% It is always \today, today,
             %  but any date may be explicitly specified

\begin{abstract}
We present the results of our QCD analysis for nonsinglet
unpolarized quark distributions and structure function
$F_2(x,Q^2)$ up to next-to-next-to-next-to leading order(N$^3$LO).
In this regards 4-loop anomalous dimension can be obtained from
the Pad\'e approximations. The analysis is based on the Jacobi
polynomials expansion of the structure function. New
parameterizations are derived for the nonsinglet quark
distributions for the kinematic wide range of $x$ and $Q^2$. Our
calculations for nonsinglet unpolarized quark distribution
functions up to N$^3$LO are in good agreement with available
theoretical models.  The higher twist contributions of
$F_2^{p,d}(x,Q^2)$ are extracted in the large $x$ region in
N$^3$LO analysis. The values of $\Lambda_{QCD}$ and
$\alpha_s(M_z^2)$ are determined.
\end{abstract}

% PACS, the Physics and Astronomy
                             % Classification Scheme.
%\keywords{Suggested keywords}%Use showkeys class option if keyword
                              %display desired
\maketitle

\section{\label{sec:Introduction}Introduction}
Structure functions in deep-inelastic scattering (DIS) and their
scale evolution are closely related to the origins of quantum
chromodynamics (QCD). DIS processes have played and still play a
very important role for our understanding of QCD and  nucleon
structure \cite{Altarelli:2009za}. In fact, DIS structure
functions have been the subject of detailed theoretical and
experimental investigations. Today, with high-precision data from
the electron proton collider, HERA, and in view of the outstanding
importance of hard scattering processes at proton–(anti)proton
colliders like the TEVATRON and the forthcoming Large Hadron
Collider (LHC) at CERN, a quantitative understanding of
deep-inelastic processes is indispensable.

To predict the rates of the various processes, a set of universal
parton distribution functions (PDF's) is required. On the other
hand all calculations of high energy processes with initial
hadrons, whether within the standard model or exploring new
physics, require PDF's as an essential input. The reliability of
these calculations, which underpins both future theoretical and
experimental progress, depends on understanding the uncertainties
of the PDF's. These distribution functions can be determined by
QCD global fits to all the available DIS and related
hard-scattering data. The QCD fits can be performed at leading
order (LO), next-to-leading order (NLO), next-to-next-to-leading
order (N$^2$LO) in the strong coupling $\alpha_s$.

The assessment of PDF's, their uncertainties and extrapolation to
the kinematics relevant for future colliders such as the LHC have
been an important challenge to high energy physics in recent
years. Over the last couple of years there has been a considerable
improvement in the precision, and in the kinematic range of the
experimental measurements for many of these processes, as well as
new types of data becoming available. In addition, there have been
valuable theoretical developments, which increase the reliability
of the global analysis. It is therefore timely, particularly in
view of the forthcoming experiments at the LHC at CERN, to perform
new global analysis which incorporate all of these improvements. A
lot of efforts and challenges have been done to obtain  PDF's for
the LHC \cite{CERN:2009} which take into account the higher order
corrections \cite{Martin:2009iq,
%Thorne:2009ky,
Thorne:2009me,Nadolsky:2008zw}.

For quantitatively reliable predictions of DIS and hard hadronic
scattering processes, perturbative QCD corrections at the N$^2$LO
and the next-to-next-to-next-to-leading order (N$^3$LO) need to be
taken into account. Based on our experience obtained in a series
of LO, NLO and N$^2$LO analysis \cite{Khorramian:2008yh} of the
nonsinglet parton distribution functions, here we extend our work
to N$^3$LO accuracy in perturbative QCD.

%The earlier studies on this topic [2] have already demonstrated
%that ...

 In this work this important problem is studied
with the help of the method of the structure function
reconstruction over their Mellin moments, which is based on the
expansion of the structure function in terms of Jacobi
polynomials. This method was developed and applied for different
QCD analyses
\cite{parisi,Barker,Krivokhizhin:1987rz,Krivokhizhin:1990ct,
Chyla:1986eb,Barker:1980wu,Kataev:1997nc,Kataev:1998ce,Kataev:1999bp,Kataev:2001kk,Khorramian:2007zz,Khorramian:2008zz,
Khorramian:2008zza,AtashbarTehrani:2009zz,Khorramian:2009zz1,Khorramian:2009zz2,Khorramian:2009zz3,Khorramian:2009zz4}.
The same method has also been applied in the polarized case in
Refs. \cite{Leader:1997kw} and \cite{Atashbar
Tehrani:2007be,Khorramian:2007gu,Khorramian:2007zza,Mirjalili:2007ep,Mirjalili:2006hf}.

In the present paper we perform a QCD analysis of the flavor
nonsinglet unpolarized deep--inelastic charged $e(\mu) p$ and
$e(\mu) d$ world data \cite{BCDMS,SLAC,NMC,H1,ZEUS} at N$^3$LO and
derived parameterizations of valence quark distributions
$xu_v(x,Q^2)$ and $xd_v(x,Q^2)$ at a starting scale $Q_0^2$
together with the QCD--scale $\Lambda_{\rm QCD}$ by using the
Jacobi polynomial expansions.  We have therefore used the 3-loop
splitting functions and Pad\'e approximations
\cite{Pade1,Pade2,Pade3,Baker,BenderOrszag} for the evolution of
nonsinglet quark distributions of hadrons.

Previous 3--loop QCD analysis were mainly performed as combined
singlet and non–singlet analysis \cite{MRST03,A02}, partly based
on preliminary, approximative expression of the 3--loop splitting
functions. Other analyses were carried out for fixed moments only
in the singlet and nonsinglet case analyzing neutrino data
\cite{SY,KAT,SYS}. First results of the nonsinglet analysis were
published in \cite{BBG04}. Very recently a 3--loop nonsinglet
analysis was also carried out in
Refs.~\cite{Khorramian:2008yh,Gluck:2006yz,Blumlein:2006be}. The
results of 4--loop QCD analysis are also reported in
\cite{Blumlein:2006be,ABKM:2009aa}. The results of the present
work are based on the Jacobi polynomials expansion of the
nonsinglet structure function.

The plan of the paper is to recall the theoretical formalism of
the QCD analysis for calculating nonsinglet sector of proton
structure function $F_2$ in Mellin-$N$ space in Sec.~II.
Section~III explains the Pad\'e approximations and 4-loop
anomalous dimensions. A description of the Jacobi polynomials
 and procedure of the QCD fit of $F_2$ data are illustrated in Sec.~IV.
The numerical results are illustrated in Sec.~V before we
summarize our findings in Sec.~VI.
\section{\label{sec:QCD Fit}Theoretical formalism of the QCD analysis}
In the common $\overline{\rm MS}$ factorization scheme the
relevant $F_2$ structure function as extracted from the DIS $ep$
process can be written as \cite{ref3,ref4,ref5,Gluck:2006pm}
\begin{eqnarray}
\label{Eq:F2-tot} x^{-1}F_2(x,Q^2) &=& x^{-1}\left(F_{2,{\rm
NS}}(x,Q^2)+ F_{2,S}(x,Q^2)+ F_{2,g}(x,Q^2)\right) \nonumber \\&=&
C_{2,{\rm NS}}(x,Q^2)\otimes q_{\rm NS}(x,Q^2)\nonumber \\&+&<e^2>
C_{2,{\rm S}}(x,Q^2)\otimes q_{\rm S}(x,Q^2)\nonumber
\\&+&<e^2>C_{2,{\rm g}}(x,Q^2)\otimes g(x,Q^2)\;,
\end{eqnarray}
here $q_i$ and $g$ represent the quarks and gluons distributions
respectively, $q_{\rm NS}$ stands for the usual flavor nonsinglet
combination and $q_S$ stand for the flavor-singlet quark
distribution, $q_S=\sum_{i=1}^{n_f}(q_i+\bar q_i)$. Also, $n_f$
denotes the number of effectively massless flavors. $<e^2>$
represents the average squared charge, and $\otimes$ denotes the
Mellin convolution which turns into a simple multiplication in
$N$-space.

The perturbative expansion of the coefficient functions can be
written as
\begin{eqnarray}
\label{Eq:WillsonExp}  C_{2,{\rm
i}}(x,\alpha_s(Q^2))=\sum_{n=0}\left(\frac{\alpha_s(Q^2)}{4\pi}\right)^n\;C_{\rm
2,i}^{(n)}(x)~.
\end{eqnarray}
In LO, $C_{\rm 2,NS}^{(0)}(x)=\delta(x)$, $C_{\rm
2,PS}^{(0)}(x)=C_{{\rm {2,}}g}^{(0)}(x)=C_{\rm 2,PS}^{(1)}(x)=0$
and the singlet-quark coefficient function is decomposed into the
nonsinglet and pure singlet contribution, $C_{\rm 2,q}^{(n)}\equiv
C_{\rm 2,S}^{(n)}=C_{\rm 2,NS}^{(n)}+C_{\rm 2,PS}^{(n)}$. The
coefficient functions $C_{\rm 2,i}^{(n)}$ up to N$^3$LO have been
given in \cite{Vermaseren:2005qc}.

 The nonsinglet structure function $F_{2,NS}(x,Q^2)$ up to N$^3$LO and for three
active (light) flavors has the representation
% with the nonsinglet contribution for three active
%(light) flavors being given by
\begin{eqnarray}
\label{Eq:F2-NS}x^{-1}F_{2,{\rm NS}}(x,Q^2)
&=&\Big[C_{2,q}^{(0)}+a_sC_{2,{\rm NS}}^{(1)}+a_s^2C_{2,{\rm
NS}}^{(2)+}+a_s^3C_{2,{\rm NS}}^{(3)+} \Big]\otimes \left[
\frac{1}{18}\, q_8^+ +\frac{1}{6}\, q_3^+\right](x,Q^2)~.
\end{eqnarray}
The flavor-singlet and gluon contributions in
Eq.~(\ref{Eq:F2-tot}) reads
\begin{eqnarray}
\label{Eq:F2-S}x^{-1}F_{2,S}(x,Q^2)&=&\frac{2}{9}  \left[
C_{2,q}^{(0)} +
a_sC_{2,q}^{(1)}+a_s^2C_{2,q}^{(2)}+a_s^3C_{2,q}^{(3)}\right]
\otimes \Sigma (x,Q^2)~;
\end{eqnarray}
\begin{eqnarray}
\label{Eq:F2-g} x^{-1} F_{2,g}(x,Q^2)&=&\frac{2}{9}
\left[a_sC_{2,g}^{(1)}+a_s^2C_{2,g}^{(2)}+a_s^3C_{2,g}^{(3)}\right]
\otimes g (x,Q^2)~.
\end{eqnarray}
The symbol $\otimes$ denotes the Mellin convolution
\begin{equation}
[A\otimes B](x)=\int_0^1 dx_1\int_0^1
dx_2\;\delta(x-x_1x_2)~A(x_1)B(x_2)\;.
\end{equation}
In Eq.~(\ref{Eq:F2-NS}) $q_3^+=u+\bar{u}-(d+\bar{d})=u_v-d_v$ and
$q_8^+= u+\bar{u}+d+\bar{d}-2(s+\bar{s}) =
u_v+d_v+2\bar{u}+2\bar{d}-4\bar{s}$, where $s=\bar{s}$. Also in
Eq.~(\ref{Eq:F2-S}),
$\Sigma(x,Q^2)\equiv\Sigma_{q=u,d,s}(q+\bar{q})=u_v+d_v+2\bar{u}+2\bar{d}+2\bar{s}$.
Notice that in the above equations
$a_s=a_s(Q^2)\equiv\alpha_s(Q^2)/4\pi$ denotes the strong coupling
constant and $C_{i,j}$ are the Wilson coefficients
\cite{Vermaseren:2005qc}.

The combinations of parton densities in the nonsinglet regime and
the valence region $x\geq0.3$ for $F_2^p$ in LO is
\begin{equation}
\frac{1}{x}\,F_2^{p}(x,Q^2) = \left[ \frac{1}{18}\, q_{{\rm
NS,}8}^+ +\frac{1}{6}\, q_{{\rm NS,}3}^+
\right](x,Q^2)+\frac{2}{9} \Sigma(x,Q^2)~,
\end{equation}
where $q_{{\rm NS,}3}^+=u_v-d_v$, $q_{{\rm NS,}8}^+=u_v+d_v$ and
$\Sigma=u_v+d_v$, since sea quarks can be neglected in the region
$x\geq0.3$. So in the $x$-space we have
\begin{eqnarray}
  F_2^{p}(x,Q^2) = \left (\frac{5}{18}\, x\, q_{{\rm NS,}8}^+
  + \frac{1}{6}\, x\, q_{{\rm NS,}3}^+\right) (x,Q^2)& = &
       \frac{4}{9}\,  x\, u_v(x,Q^2)+\frac{1}{9}\, x\, d_v(x,Q^2)~.
\end{eqnarray}
In the above region the combinations of parton densities for
$F_2^d$ are also given by
\begin{eqnarray}
  F_2^{d}(x,Q^2) = \left (\frac{5}{18}\, x\, q_{{\rm NS,}8}^+\right) (x,Q^2)& = &
      \frac{5}{18}\, x(u_v+d_v)(x,Q^2)~,
\end{eqnarray}
where $q_{{\rm NS,}3}^+=u_v-d_v$ and $F_2^d=(F_2^p+F_2^n)/2$ if we
ignore the nuclear effects here. It is important to stress that
the shadowing effect as a nuclear effect may affect our analysis.
The shadowing effect \cite{Barone:1992ej,Kwiecinski:1987tb}
arising from the gluon recombination and  in the small-$x$ region,
the competitive mechanism of nuclear shadowing takes place. It
also depends on the size of the nucleons. According to this effect
we have $F_2^d=(F_2^p+F_2^n)/2+\delta F_2^d$. To obtain the
$\delta F_2^d$ we need to know the generalized vector meson
dominance (VMD) and parton mechanism at low and large values of
$Q^2$ respectively. We found that the value of $\delta F_2^d$ is
important but in low values of $x$. For example this correction
value at $Q^2$=10 GeV$^2$ and for $x>0.1$ is too small ($\sim
10^{-4}$).
 So in the valence region of this analysis, this effect
is negligible in large $x$ and we can use the
$F_2^d=(F_2^p+F_2^n)/2$ approximately.

 In the region $x
\leq 0.3$ for the difference of the proton and deuteron data we
use
\begin{eqnarray}
F_2^{NS}(x,Q^2)&\equiv& 2 (F_2^{p}-F_2^{d})(x,Q^2)\nonumber \\
&=&\frac{1}{3}\, x\, q_{{\rm NS,}3}^+(x,Q^2)= \frac{1}{3}\,
x(u_v-d_v)(x,Q^2)
  +\frac{2}{3}\, x(\bar{u}-\bar{d})(x,Q^2)~,
\end{eqnarray}
where now $q_{{\rm NS,}3}^+ = u_v-d_v+2(\bar{u}-\bar{d})$ since
sea quarks cannot be neglected for $x$ smaller than about 0.3.

The first clear evidence for the flavor asymmetry combination of
light parton distributions $x(\overline{d}-\overline{u})$ in
nature came from the analysis of NMC at CERN to study of the
Gottfried sum rule \cite{Amaudruz:1991at}. In our calculation we
supposed the $\bar{d}-\bar{u}$ distribution
\cite{Gluck:2006yz,Blumlein:2006be,ref19,Blumlein:2004pr}
%eq.(4)
\begin{equation}
x(\bar{d}-\bar{u})(x,Q_0^2) = 1.195 x^{1.24}(1-x)^{9.10}
  (1 + 14.05x - 45.52 x^2)~,
  \label{Asy}
\end{equation}
at $Q_0^2=4$ GeV$^2$ which gives a good description of the
Drell-Yan dimuon production data \cite{E866}. In this analysis,
like other analyses
\cite{Khorramian:2008yh,Khorramian:2007zz,Gluck:2006yz,Blumlein:2006be,ref19,Blumlein:2004pr},
we used the above distribution for considering the symmetry
breaking of sea quarks. Although, in fact, this parametrization
plays a marginal role in our analysis, in order to find the impact
effect of this distribution, which is essentially used in the
paper, it is desirable to study the QCD fits by varying this
distribution with another asymmetry sea quark distribution which
is derived in other analyses. In Sec. VI we will discuss our
outputs when we change the above sea distribution.

Now these results in the physical region $0<x\leq 1$ can transform
to Mellin-$N$ space  by using the Mellin transform to obtain the
moments of the structure function as $\frac{1}{x}F_2^k$,
\begin{equation}
F_2^k(N,Q^2)
\equiv{\bf{M}}[F_2^k,N]=\int_0^1dx~x^{N-1}\frac{1}{x}F_2^k(x,Q^2)~,
  \label{Asy}
\end{equation}
here $k$ denotes the three above cases, i.e. $k=p,d, NS$.  One of
the advantages of Mellin-space calculations is the fact that the
Mellin transform of a convolution of functions in
Eqs.~(\ref{Eq:F2-NS},\ref{Eq:F2-S},\ref{Eq:F2-g}) reduces to a
simple product

\begin{equation}
{\bf{M}}[A\otimes B,N]={\bf{M}}[A,N]{\bf{M}}[B,N]=A(N)B(N)
  \label{Momconv}
\end{equation}

By using the solution of the nonsinglet evolution equation for the
parton densities to 4$-$ loop order, the nonsinglet structure
functions are given by  \cite{Blumlein:2006be}
\begin{eqnarray}
F_2^k(N,Q^2)&=&\left(1+a_s\;C_{2,{\rm
NS}}^{(1)}(N)+a_s^2\;C_{2,{\rm NS}}^{(2)}(N)+a_s^3\;C_{2,{\rm
NS}}^{(3)}(N)\right) F_2^k(N,Q_0^2)
 \nonumber
\\
&&\times\left(\frac{a_s}{a_0}\right)^{-\hat{P}_0(N)/{\beta_0}}\Biggl\{1
- \frac{1}{\beta_0} (a_s - a_0) \left[\hat{P}_1^+(N)
- \frac{\beta_1}{\beta_0} \hat{P}_0(N) \right] \nonumber\\
& & - \frac{1}{2 \beta_0}\left(a_s^2 - a_0^2\right)
\left[\hat{P}_2^+(N) - \frac{\beta_1}{\beta_0} \hat{P}_1^+(N) +
\left( \frac{\beta_1^2}{\beta_0^2} - \frac{\beta_2}{\beta_0}
\right) \hat{P}_0(N)   \right] \nonumber\\ & & + \frac{1}{2
\beta_0^2} \left(a_s - a_0\right)^2 \left(\hat{P}_1^+(N) -
\frac{\beta_1}{\beta_0} \hat{P}_0(N) \right)^2 \nonumber\\ & &
%---
- \frac{1}{3 \beta_0} \left(a_s^3 - a_0^3\right)
\Biggl[\hat{P}_3^+(N) - \frac{\beta_1}{\beta_0} \hat{P}_2^+(N) +
\left(\frac{\beta_1^2}{\beta_0^2} - \frac{\beta_2}{\beta_0}\right)
\hat{P}_1^+(N) \nonumber\\ & & +\left(\frac{\beta_1^3}{\beta_0^3}
-2 \frac{\beta_1 \beta_2}{\beta_0^2} + \frac{\beta_3}{\beta_0}
\right) \hat{P}_0(N)  \Biggr] \nonumber\\ & & + \frac{1}{2
\beta_0^2} \left(a_s-a_0\right)\left(a_0^2 - a_s^2\right)
\left(\hat{P}_1^+(N)-\frac{\beta_1}{\beta_0} \hat{P}_0(N) \right)
\nonumber\\ & & \times \left[\hat{P}_2(N) -
\frac{\beta_1}{\beta_0} \hat{P}_1(N) -
\left(\frac{\beta_1^2}{\beta_0^2} - \frac{\beta_2}{\beta_0}
\right) \hat{P}_0(N)
 \right]
\nonumber\\ && - \frac{1}{6 \beta_0^3} \left(a_s-a_0\right)^3
\left(\hat{P}_1^+(N)-\frac{\beta_1}{\beta_0} \hat{P}_0(N)
\right)^3 \Bigg\}~.
\end{eqnarray}
Here $a_s(=\alpha_s/4\pi)$ and  $a_0$ denotes the strong coupling
constant in the scale of $Q^2$ and $Q_0^2$ respectively. $k=p,d$
and $NS$ also denotes the three above cases, i.e. proton, deuteron
and nonsinglet structure function. $C_{2,NS}^{(m)}(N)$ are the
nonsinglet Wilson coefficients in ${\it{O}}(a_s^m)$ which can be
found in \cite{FP,NS2,Vermaseren:2005qc} and $\hat{P}_m$ denote
also the Mellin transforms of the $(m+1)-$ loop splitting
functions.

\section{\label{sec:QCD Fit}Pad\'e approximations and 4-loop anomalous dimensions}

In spite of the unknown 4-loop anomalous dimensions, one can
obtain the nonsinglet parton distributions and $\Lambda_{QCD}$ by
estimating uncalculated fourth-order corrections to the nonsinglet
anomalous dimension. On the other hand the 3--loop Wilson
coefficients are known \cite{Vermaseren:2005qc} and now it is
possible to know, which effect has the 4-loop anomalous dimension
if compared to the Wilson coefficient. In this case the 4-loop
anomalous dimension may be obtained from Pad\'e approximations.

Pad\'e approximations have proved to be useful in many physical
applications. Pad\'e approximations may be used either to predict
the next term in some perturbative series, called a Pad\'e
approximation prediction, or to estimate the sum of the entire
series, called Pad\'e summation.

For this purpose we use the Pad\'e approximations of the
perturbative series, discussed in detail for QCD, e.g., in
Refs.~\cite {Pade1,Pade2,Pade3}. Pad\'e approximations
\cite{Baker,BenderOrszag} are rational functions chosen to equal
the perturbative series to the order calculated:
\begin{eqnarray}
 [{\cal N}/{\cal M}]=\frac{a_0 + a_1x + ...
+a_{\cal N}x^{\cal N}}{1 + b_1x + ... + b_{\cal M}x^{\cal M}}~,
\label{PadeDef}
\end{eqnarray}
to the series
\begin{eqnarray}
 \label{Spade}
 S=S_0+S_1x+...+S_{{\cal N}+{\cal M}}x^{{{\cal N}+{\cal M}}}~,
\end{eqnarray}
where we set
\begin{eqnarray}
 \label{formII pade}
[{\cal N}/{\cal M}] = S + {\cal O}(x^{{\cal N}+{\cal M}+1})~,
\phantom{aaa} \nonumber
\end{eqnarray}
and write an equation for the coefficients of each power of $x$.
To continue, let's go to Mellin-$N$ space.

A generic QCD anomalous dimension expansion in term of $a_s$ then
may be written in the form
\begin{eqnarray}
 \label{Anom pade}
\gamma(N)=\sum_{l=0}^{\infty}a_s^{l+1}\gamma^{(l)}(N)~.
\phantom{aaa}
\end{eqnarray}
In Mellin-$N$ space and by using this approach we can replace
$\gamma(N)$ by a rational function in $a_s$
\cite{Vermaseren:2005qc},
\begin{eqnarray} \label{eq:pade}
 \widetilde{\gamma}^{\:[\cal{N}/\cal{M}]}(N) \equiv [{\cal{N}/\cal{M}}](N) =
 \frac{p_{0} + a_s p_{1}(N) + \ldots + a_s^{\cal N} p_{\cal N}(N)}
 {1 + a_s q_{1}(N) + \ldots + a_s^{\cal M} q_{\cal M}(N)} \:\: .
\end{eqnarray}
Here ${\cal M} \,\geq\, 1$ and ${\cal N} + {\cal M} \, = \, n$,
where $n$ stands for the maximal order in $a_s$ at which the
expansion coefficients $\gamma^{(n)}(N)$ have been determined from
an exact calculation. The functions $p_i(N)$ and $q_{\!j}(N)$ are
determined from these known coefficients by expanding
Eq.~(\ref{eq:pade}) in powers of $a_s$. This expansion then also
provides the $[\cal{N}/\cal{M}]$ Pad\'e approximate for the
($n\!+\!1$)-th order quantities $\gamma^{(n\!+\!1)}$.

In this way it is easy to obtain the following results for ${\cal
M}={\cal N}=1$ and for  ${\cal M}=0,{\cal N}=2$
\begin{eqnarray}
 \label{pade11}
 \widetilde{\gamma}^{\:[1/1]}(N)&\equiv&[1/1](N)=\frac{\gamma^{(2)^2}(N)}{\gamma^{(1)}(N)}~,\nonumber
 \\
 \widetilde{\gamma}^{\:[0/2]}(N)&\equiv&[0/2](N)=\frac{2\gamma^{(1)}(N)\gamma^{(2)}(N)}
 {\gamma^{(0)}(N)}-\frac{\gamma^{(1)^3}(N)}{\gamma^{(0)^2}(N)}~.
\end{eqnarray}

\begin{figure}
\vspace{2 cm}
\includegraphics[angle=0, width=8.0cm]{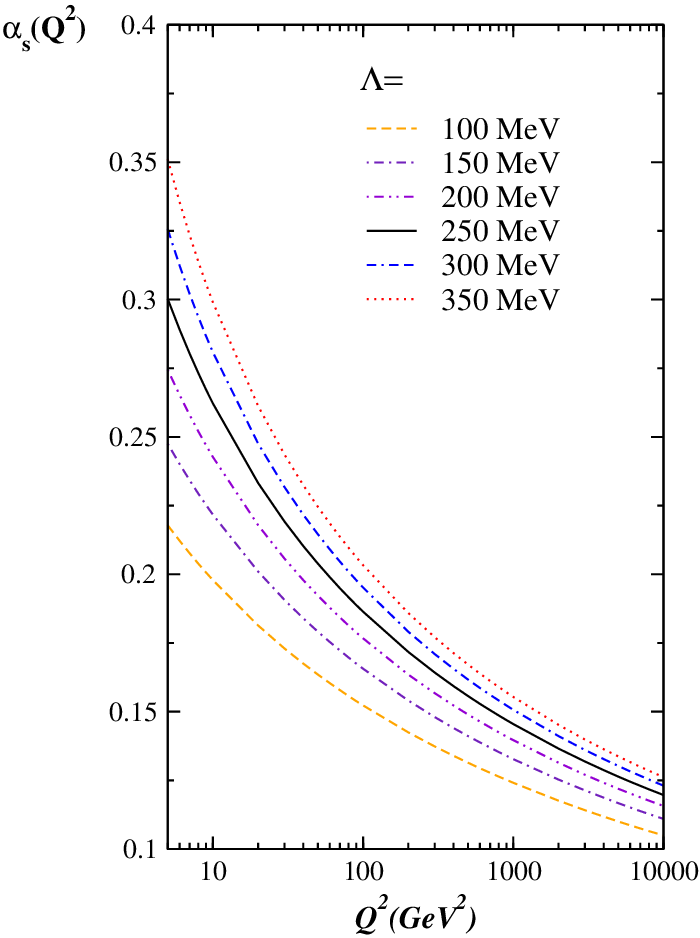}% Here is how to import EPS art
\caption{\label{fig:0}{\sf The strong running of $\alpha_s(Q^2)$,
according to Eq.~\ref{as-exp1}, in 4- loop approximation and for
different values of $\Lambda$. }}
\end{figure}

The strong coupling constant $a_{s}$ plays a more central role in
the present paper to the evolution of parton densities. At
$N^{m}LO$ the scale dependence of $a_{s}$ is given by
\begin{eqnarray}
 \label{as-eqn}
  \frac{d\, a_{s}}{d \ln Q^2} \; = \; \beta_{N^mLO}(a_{s})
  \; = \; - \sum_{k=0}^m \, a_{s}^{k+2} \,\beta_k \;.
\end{eqnarray}
The expansion coefficients $\beta_k$ of the $\beta$-function of
QCD are known up to $k=3$, i.e., N$^3$LO
\cite{Tarasov:1980au,Larin:1993tp}
\begin{eqnarray}
 \label{beta-exp}
  \beta_0 &=& 11-2/3~n_f\;,
  \nonumber \\
  \beta_1 &=& 102-38/3~n_f\;,
  \nonumber \\
  \beta_2 &=& 2857/2-5033/18~n_f+325/54~n_f^2\;,
  \nonumber \\
  \beta_3 &=& 29243.0 - \: 6946.30~n_f + 405.089~n_f^2
                  + 1093/729~n_f^3~,
\end{eqnarray}
here $n_f$ stands for the number of effectively massless quark
flavors and $\beta_k$ denote the coefficients of the usual
four-dimensional $\overline{MS}$ beta function of QCD. In complete
4-loop approximation and using the $\Lambda$-parametrization, the
running coupling is given by \cite{Vogt:2004ns,Chetyrkin:1997sg}:
\begin{eqnarray}
\label{as-exp1}
 a_s(Q^2)
 &=& \frac{1}{\beta_0{L_{\Lambda}}}  -
    \frac{1}{(\beta_0{L_{\Lambda}})^2}~b_1 \ln {L_{\Lambda}} \nonumber\\  &+&
    \frac{1}{(\beta_0{L_{\Lambda}})^3} \left[b_1^2 \left(\ln^2 {L_{\Lambda}}-\ln {L_{\Lambda}}-1
    \right) + b_2\right] \nonumber\\
 &+& \frac{1}{(\beta_0{L_{\Lambda}})^4}\left[b_1^3 \left(-\ln^3 {L_{\Lambda}} +
    \frac{5}{2} \ln^2{L_{\Lambda}} +2 \ln{L_{\Lambda}}-\frac{1}{2}\right) - 3b_1 b_2\ln{L_{\Lambda}}
    +\frac{b_3}{2} \,\right]~,
\end{eqnarray}
where $L_{\Lambda}\equiv ln (Q^2/\Lambda^2)$, $b_k\equiv \beta_k
/\beta_0$, and $\Lambda$ is the QCD scale parameter. The first
line of Eq.~(\ref{as-exp1}) includes the 1- and the 2-loop
coefficients, the second line is the 3-loop and the third line
denotes the 4-loop correction. Equation~(\ref{as-exp1}) solves the
evolution equation (\ref{as-eqn}) only up to higher orders in
$1/L_{\Lambda}$. The functional form of $\alpha_s(Q^2)$, in 4-loop
approximation and for 6 different values of $\Lambda$, is
displayed in Fig.~\ref{fig:0}. The slope and dependence on the
actual value of $\Lambda$ is especially pronounced at small $Q^2$,
while at large $Q^2$ both the energy dependence and the dependence
on $\Lambda$ becomes increasingly feeble. To be able to compare
with other measurements of $\Lambda$ we adopt the matching of
flavor thresholds at $Q^2=m_c^2$ and $Q^2=m_b^2$ with $m_c=1.5$
GeV and $m_b=4.5$ GeV as described in \cite{BAR,R1998}.

\section{Jacobi polynomials and the procedure of QCD fits}

One of the simplest and fastest possibilities in the structure
function reconstruction from the QCD predictions for its Mellin
moments is Jacobi polynomials expansion. The Jacobi polynomials
are especially suitable for this purpose since they allow one to
factor out an essential part of the $x$-dependence of  structure
function into the weight function \cite{parisi}.

According to this method, one can relate the $F_2$ structure
function with its Mellin moments
\begin{eqnarray} F_{2}^{~k,N_{max}}(x,Q^2)&=&x^{\beta}(1-x)^{\alpha}
\sum_{n=0}^{N_{max}}\Theta_n ^{\alpha,
\beta}(x)\sum_{j=0}^{n}c_{j}^{(n)}{(\alpha ,\beta )}
F_{2}^k(j+2,Q^2), \label{eg1Jacob} \end{eqnarray} where $N_{max}$
is the number of polynomials, $k$ denotes the three  cases, i.e.
$k=p,d, NS$.  Jacobi polynomials of order $n$
\cite{Parisi:1978jv}, $\Theta_n ^{\alpha, \beta}(x)$, satisfy the
orthogonality condition with the weight function $w^{\alpha
\beta}=x^{\beta}(1-x)^{\alpha}$
\begin{equation}
\int_{0}^{1}dx\;w^{\alpha \beta} \Theta_{k} ^{\alpha , \beta}(x)
\Theta_{l} ^{\alpha , \beta}(x)=\delta_{k,l}\ .\label{e8}
\end{equation}
In the above, $c_{j}^{(n)}{(\alpha ,\beta )}$ are the coefficients
expressed through $\Gamma$-functions and satisfying the
orthogonality relation in Eq.~(\ref{e8}) and $F_{2}(j+2,Q^2)$ are
the moments determined in the previous section. $N_{max}$,
$\alpha$ and $\beta$  have to be chosen so as to achieve the
fastest convergence of the series on the right-hand side of
Eq.~(\ref{eg1Jacob}) and to reconstruct $F_2$ with the required
accuracy. In our analysis we use $N_{max}=9$, $\alpha=3.0$ and
$\beta=0.5$. The same method has been applied to calculate the
nonsinglet structure function $xF_3$ from their moments
\cite{Kataev:1997nc,Kataev:1998ce, Kataev:1999bp,Kataev:2001kk}
and for polarized structure function $xg_1$ \cite{Atashbar
Tehrani:2007be,Leader:1997kw,Khorramian:2007gu}. Obviously the
$Q^2$-dependence of the polarized structure function is defined by
the $Q^2$-dependence of the moments.

The evolution equations allow one to calculate the
$Q^2$-dependence of the parton distributions  provided  at a
certain reference point $Q_0^2$. These distributions are usually
parameterized on the basis of plausible theoretical assumptions
concerning their behavior near the end points $x=0,1$.

%\section{\label{sec:QCD Fit}The  Procedure of the QCD Fits of  $F_2$ Data}
%%%%%%%%%%%%%%%%%%%%%%%%%%%%%%%%%%%%%%%%%%%%%%%%%%%%%%%%%%%%%%%%%%%%%%%

In the present analysis we choose the following parametrization
for the valence quark densities in the input scale of $Q^2_0=4$
GeV$^2$
\begin{eqnarray}
\label{equ:param} x q_v(x,Q^2_0) = {{\cal N}}_q~x^{a_q}(1-x)^{b_q}
(1 + c_q \sqrt{x} + d_q~x)~,
\end{eqnarray}
where $q=u,d$ and the normalization factors ${{\cal N}}_u$ and
${{\cal N}}_d$ are fixed by $\int_0^1 u_v dx=2$ and $\int_0^1 d_v
dx=1$, respectively.  By QCD fits of the world data for
$F_2^{p,d}$, we can extract valence quark densities using the
Jacobi polynomials method. For the nonsinglet QCD analysis
presented in this paper we use the structure function data
measured in charged lepton-proton and  deuteron deep-inelastic
scattering. The experiments contributing to the statistics are
BCDMS~\cite{BCDMS}, SLAC~\cite{SLAC}, NMC~\cite{NMC},
H1~\cite{H1}, and ZEUS~\cite{ZEUS}. In our QCD analysis we use
three data samples~: $F_2^p(x,Q^2)$, $F_2^d(x,Q^2)$ in the
nonsinglet regime and the valence quark region $x \geq 0.3$ and
$F_2^{NS} = 2 (F_2^p - F_2^d)$ in the region $x < 0.3$.

The valence quark region may be parameterized by the nonsinglet
combinations of  parton distributions, which are expressed through
the parton distributions of valence quarks. Only data with $Q^2 >
4~\GeV^2$ were included in the analysis and a cut in the hadronic
mass of $W^2\equiv (\frac{1}{x}-1)\, Q^2+m_{\rm N}^2
> 12.5~\GeV^2$ was applied in order to widely eliminate higher
twist (HT) effects from the data samples. After these cuts we are
left with 762 data points, 322 for $F_2^p$, 232 for $F_2^d$, and
208 for $F_2^{NS}$. By considering the additional cuts on the
BCDMS ($y> 0.35$) and on the NMC data($Q^2 > 8$ GeV$^2$) the total
number of data points available for the analysis reduce from 762
to 551, because we have 227 data points for $F_2^p$, 159 for
$F_2^d$, and 165 for $F_2^{NS}$.

For data used in the global analysis, most experiments combine
various systematic errors into one effective error for each data
point, along with the statistical error. In addition, the fully
correlated normalization error of the experiment is usually
specified separately. For this reason, it is natural to adopt the
following definition for the effective $\chi^2$
\cite{Stump:2001gu,Khorramian:2008yh}
\begin{eqnarray}
\chi _{\mathrm{global}}^{2} &=&
\sum_{n} w_{n} \chi _{n}^{2}\;,\qquad (n\;%
\mbox{labels the different experiments})\nonumber
\label{eq:Chi2global}
\\
\chi _{n}^{2} &=&\left(\frac{1-{\cal N}_{n}}{\Delta{\cal
N}_{n}}\right)^{2} +\sum_{i}\left( \frac{{\cal
N}_{n}F_{2,i}^{data}-F_{2,i}^{theor}}{{\cal N}_{n}\Delta
F_{2,i}^{data}} \right)^{2}\;. \label{eq:Chi2n}
\end{eqnarray}

For the $n^{\mathrm{th}}$ experiment, $F_{2,i}^{data}$, $\Delta
F_{2,i}^{data}$, and $%
F_{2,i}^{theor}$ denote the data value, measurement uncertainty
(statistical and systematic combined) and theoretical value for
the $i^{\mathrm{th}}$ data point. ${\Delta{\cal N}_{n}}$ is the
experimental normalization uncertainty and ${\cal N}_{n}$ is an
overall normalization factor for the data of experiment $n$. The
factor $w_{n}$ is a possible weighting factor (with default value
1).  However, we allowed for a relative normalization shift ${\cal
N}_{n}$ between the different data sets within the normalization
uncertainties ${\Delta{\cal N}_{n}}$ quoted by the experiments.
For example the normalization uncertainty of the NMC(combined)
data is estimated to be 2.5\%.
 The normalization shifts ${\cal N}_{n}$ were fitted
once and then kept fixed.

Now the sums in $\chi _{\mathrm{global}}^{2}$ run over all data
sets and in each data set over all data points. The minimization
of the  above $\chi^2$ value to determine the best parametrization
of the unpolarized parton distributions is done using the program
{\tt MINUIT} \cite{MINUIT}.

The  one $\sigma$ error for the parton density $xq_v$ as given by
Gaussian error propagation is \cite{Blumlein:2006be}
\begin{eqnarray}
\sigma( xq_v(x))^2 =  \sum_{i=1}^{n_p}\sum_{j=1}^{n_p}
                \left( \frac{\partial xq_v}{\partial p_i}\right)
                \left(\frac{\partial xq_v}{\partial p_j} \right)
                \textrm{cov}(p_i,p_j)~,
\end{eqnarray}
 where the sum runs over all fitted parameters. The functions
$\partial xq_v /
\partial p_i$ are the derivatives of $xq_v$ with respect to the fit parameter $p_i$,
and $\textrm{cov}(p_i,p_j)$ are the elements of the covariance
matrix. The derivatives $\partial xq_v / \partial p_i$ can be
calculated analytically at the input scale $Q_0^2$. Their values
at $Q^2$ are given by evolution which is performed in {Mellin-$N$}
space.

Now we need to discuss the derivatives in Mellin-$N$ space a bit
further. The Mellin-$N$ moment for complex values of $N$
calculated at the input scale $Q_0^2$ for the parton density
parameterized as in Eq.~(\ref{equ:param}) is given by
\begin{eqnarray}
q_{v}(N,a_{q},b_{q},c_{q},d_{q})&=&{\cal{N}}_{q}~{\bf{M}}(n,a_{q},b_{q},c_{q},d_{q})~,
\end{eqnarray}
with the normalization constant
\begin{equation}
{\cal{N}}_{q}=
\frac{C_{q_v}}{{\bf{M}}(1,a_{q},b_{q},c_{q},d_{q})}~.
\end{equation}
Here $C_{q_v}$ is the respective number of valence quarks, i.e.
$C_{u_v}$=2 and $C_{d_v}$=1. In the above
${\bf{M}}(n,a_{q},b_{q},c_{q},d_{q})$  is given by
\begin{eqnarray}
{\bf{M}}(n,a_{q},b_{q},c_{q},d_{q})=B[a_{q}+n-1,b_{q}+1]+c_{q}B[a+n+1/2,b+1]+d_{u}B[a_{q}+n,b_{q}+1]~,\nonumber \\
\end{eqnarray}
where $B[a,b]$ denotes the Euler beta function for complex
arguments. The general form of the derivative of the Mellin moment
$q_{v}$ with respect to the parameter $p$ is given by
\begin{equation}
\frac{\partial q_{v}(N,p)}{\partial p}={\bf{M}}(n,p)\frac{\partial {\cal{N}}_{q}}{%
\partial p}+{\cal{N}}_{q}\frac{\partial {\bf{M}}(n,p)}{\partial
p}~.
\end{equation}
In this analysis only the parameters $a_q$ and  $b_q$ have been
fitted for both the $xu_v$ and $xd_v$ parametrization while the
other parameters involved are kept fixed  after a first
minimization in the MINUIT program, since their errors turned out
to be rather large compared to the central values. Here we want to
show the derivatives $u_v$ and $d_v$ parton densities with respect
to parameter $a_q$ and $b_q$. For example:

\begin{eqnarray}
f(n,a_q) &\equiv&\frac{\partial {\bf{M}}(n,a_q)}{\partial a_q}=B[a_{q}+n-1,b_{q}+1](%
\psi [a_{q}+n-1]-\psi [a_{q}+b_{q}+n])+ \nonumber\\
&&c_{q}B[a_{q}+n-1/2,b_{q}+1](\psi [a_{q}+n-1/2]-\psi
[a+b+n+1/2])+
\nonumber \\
&&d_{q}B[a_{q}+n,b_{q}+1](\psi [a_{q}+n]-\psi [a_{q}+b_{q}+n+1])~,
\end{eqnarray}

\begin{eqnarray}
f(n,b_q) &\equiv&\frac{\partial {\bf{M}}(n,b_q)}{\partial b_{q}}=B[a_{q}+n-1,b_{q}+1](%
\psi [b_{q}+1]-\psi [a_{q}+b_{q}+n])+  \nonumber\\
&&c_{q}B[a_{q}+n-1/2,b_{q}+1](\psi [1+b_{q}]-\psi
[a_{q}+b_{q}+n+1/2])+
\nonumber \\
&&d_{q}B[a_{q}+n,b_{q}+1](\psi [b_{q}+1]-\psi [a_{q}+b_{q}+n+1])~,
\end{eqnarray}
and now we can reach the below derivatives for $u_v(N)$ and
$d_v(N)$ with respect to parameters $a_q$ and $b_q$
\begin{equation}
\frac{\partial q_{v}(N,p)}{\partial p}%
={\cal{N}}_{q}\left(f(n,p)-f(1,p){\bf{M}}(n,p)/{\bf{M}}(1,p)\right)~,
\end{equation}
also $\psi [n]={d~ln~\Gamma(n)}/{dn}$ is Euler's  $\psi$-function.

To obtain the error calculation of the structure functions
$F_2^p$,  $F_2^d$ , and  $F_2^{NS}$ the relevant gradients of the
PDF's in Mellin space have to be multiplied with the corresponding
Wilson coefficients. This yields the errors as far as the QCD
parameter $\Lambda$ is fixed and regarded uncorrelated. The error
calculation for a variable $\Lambda$  is done numerically due to
the non–linear relation and required iterative treatment in the
calculation of $\alpha_s(Q^2,\Lambda)$
\cite{Khorramian:2008yh,Blumlein:2006be}.

\section{Results}
\begin{figure}
\vspace{2 cm}
\includegraphics[angle=0, width=12.0cm]{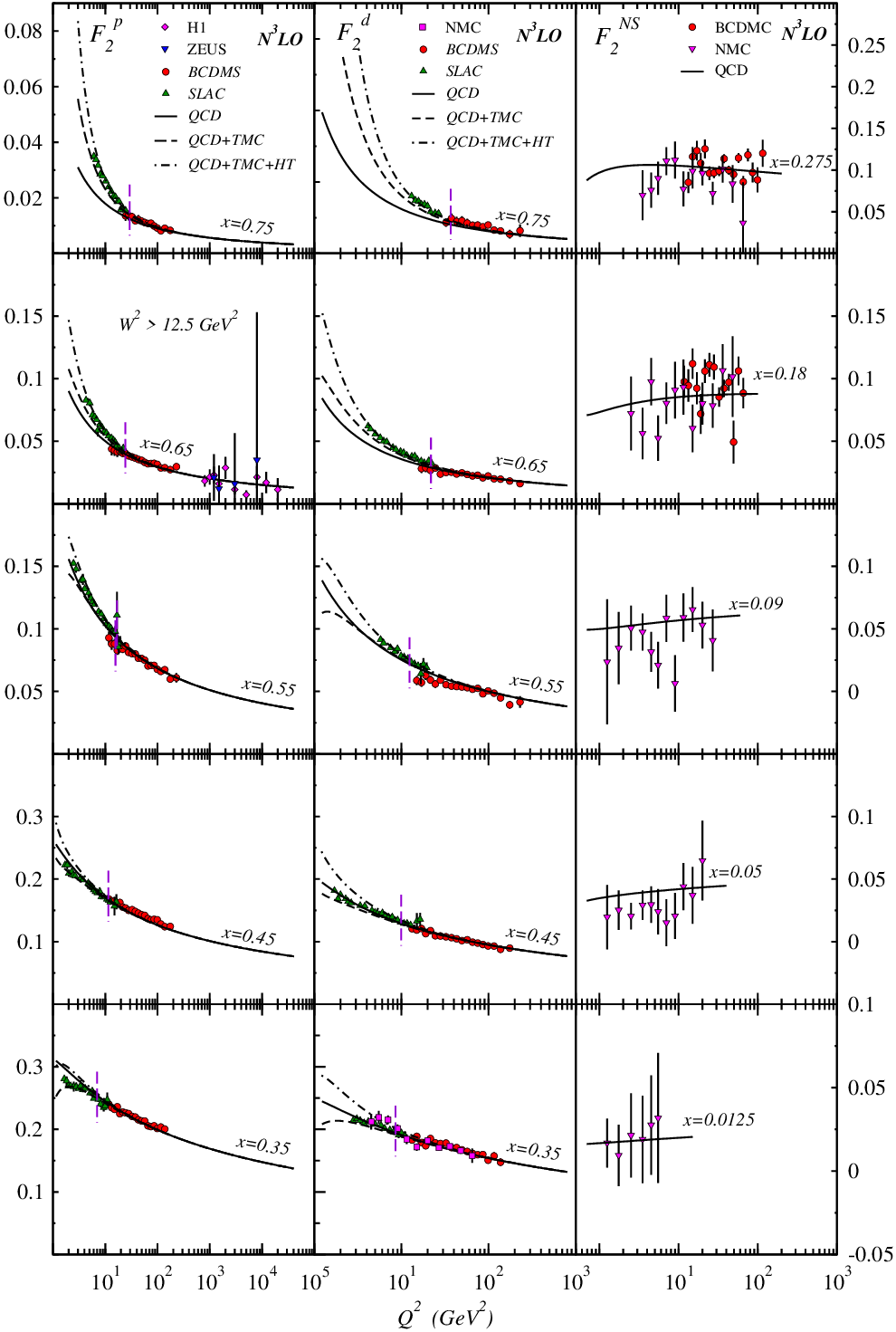}% Here is how to import EPS art
\caption{\label{fig:1}{\sf The structure functions $F_2^p$,
$F_2^d$, and $F_2^{NS}$  as a function of $Q^2$ in intervals of
$x$. Shown are the Pad\'e [1/1] QCD fits in N$^3$LO (solid line)
and the contributions from target mass corrections (dashed line)
and higher twist (dash--dotted line). The vertical dashed line
indicates the regions with $W^2 > 12.5$ GeV$^2$.
%The vertical dashed line
%indicate the regions with $W^2 > 12.5~{\GeV^2}$.
}}
\end{figure}

%%%%%%%%%%%%%%%%%%%%%%%%%%%%%%%%%%%%%%%%%%%%%%%%%%%%%%%%%%%%%%%%%%%%%%%%
In the QCD analysis of the present paper we used three data sets:
the structure functions $F_2^{p}(x,Q^2)$  and $F_2^{d}(x,Q^2)$ in
the region  of $x \geq 0.3$ and the combination of these structure
functions $F_2^{\rm NS}(x,Q^2)$ in the region  of $x < 0.3$~.
Notice that we take into account the cuts $Q^2>4$ GeV$^2$,
$W^2>12.5$ GeV$^2$ for our QCD fits to determine some unknown
parameters. In Fig.(\ref{fig:1}) the proton, deuteron and
nonsinglet data for $F_2^p(x,Q^2)$, $F_2^d(x,Q^2)$ and
$F_2^{NS}(x,Q^2)$ are shown in the nonsinglet regime and the
valence quark region $x \geq 0.3$ indicating the above cuts by a
vertical dashed line. The solid lines correspond to the N$^3$LO
QCD fit. Now, it is possible to take into account the target mass
effects in our calculations. The perturbative form of the moments
is derived under the assumption that the mass of the target hadron
is zero (in the limit $Q^{2}\rightarrow\infty$). At intermediate
and low $Q^{2}$ this assumption will begin to break down and the
moments will be subject to potentially significant power
corrections, of order ${{\cal O}}~(m_{N}^{2}/Q^{2})$, where $m_N$
is the mass of the nucleon. These are known as target mass
corrections (TMCs) and when included, the moments of flavor
nonsinglet structure function have the form
\cite{Georgi:1976ve,Gluck:2006yz}
\begin{eqnarray}
\label{equ:TMC} F_{2,{\rm TMC}}^{k}(n,Q^2) & \equiv &
  \int_0^1 x^{n-1} \frac{1}{x}F_{2,{\rm TMC}}^{k}(x,Q^2)\, dx  \nonumber\\
& = &F_{2}^{
k}(n,Q^2)+\frac{n(n-1)}{n+2}\left(\frac{m_N^2}{Q^2}\right)\,F_{2}^{
k}(n+2,Q^2)\nonumber \\
&&+\frac{(n+2)(n+1)n(n-1)}{2(n+4)(n+3)}\left(\frac{m_N^2}{Q^2}\right)^2\,F_{2}^{
k}(n+4,Q^2)+ {\cal{O}}\left( \frac{m_N^2}{Q^2}\right)^3~,
\end{eqnarray}
where higher powers than $(m_{\rm N}^2/Q^2)^2$ are negligible for
the relevant $x <0.8$ region. By inserting Eq.~(\ref{equ:TMC}) in
Eq.~(\ref{eg1Jacob}) we have
\begin{eqnarray}
F_{2}^{~k, N_{max}}(x,Q^2)&=&x^{\beta}(1-x)^{\alpha}
\sum_{n=0}^{N_{max}}\Theta_n ^{\alpha,
\beta}(x)\times\sum_{j=0}^{n}c_{j}^{(n)}{(\alpha ,\beta )}
F_{2,{\rm TMC}}^{k}(j+2,Q^2)\;, \label{eg1JacobTMC} \end{eqnarray}
where $F_{2,{\rm TMC}}^{k}(j+2,Q^2)$ are the moments determined by
Eq.~(\ref{equ:TMC}). In Fig.(\ref{fig:1})  the dashed lines
correspond to the N$^3$LO QCD fit adding target mass corrections.

Despite the kinematic cuts ($Q^2\geq 4$ GeV$^2$, $\, W^2\equiv
(\frac{1}{x}-1)\, Q^2+m_{\rm N}^2\geq 12.5$ GeV$^2$) used for our
analysis, we also take into account higher twist corrections to
$F_2^{p}(x,Q^2)$ and $F_2^{d}(x,Q^2)$ in the kinematic region $Q^2
\geq 4 \GeV^2, 4<W^2 < 12.5 \GeV^2$ in order to learn whether
nonperturbative effects may still contaminate our perturbative
analysis. For this purpose we extrapolate the QCD fit results
obtained for $W^2 \geq 12.5 \GeV^2$ to the region $Q^2 \geq 4
\GeV^2, 4<W^2 < 12.5 \GeV^2$ and from the difference between data
and theory, applying target mass corrections in addition. Now by
considering higher twist correction
%-----------------------------------------------------------------------
\begin{eqnarray}
\label{equ:HTC} F_2^{\rm exp}(x,Q^2) = O_{\rm TMC}[F_2^{\rm
HT}(x,Q^2)] \cdot \left( 1 +
\frac{h(x,Q^2)}{Q^2[\GeV^2]}\right)\;,
\end{eqnarray}
%-----------------------------------------------------------------------
the higher twist coefficient can be extract. Here the operation
$O_{\rm TMC}[...]$ denotes taking the target mass corrections of
the twist--2 contributions to the respective structure function.
The coefficients $h(x,Q^2)$ are determined in bins of $x$ and
$Q^2$ and are then averaged over $Q^2$.  We extrapolate our QCD
fits to the region $12.5 \GeV^2 \geq W^2 \geq 4 \GeV^2$ in
Fig.(\ref{fig:1}). The dash-dotted lines in this figure
  correspond to the N$^3$LO QCD fit adding target mass and higher twist corrections.
 There, at higher values of $x$ a clear gap between the
data and the QCD fit is seen.

%%%%%%%%%%%%%%%%%%%%%%%%%%%%%%%%%%%%%%%%%%%%%%%%%%%%%%%%%%%%%%%%%%%%%%%%
%\

\begin{table*}
\begin{tabular}{|c|c|c|c|c|c|}
\hline \hline
           &              &     NLO     &     N$^{2}$LO   &    N$^{3}$LO Pad\'e~[1/1]   &    N$^{3}$LO Pad\'e~[0/2]  \\
\hline $u_v$      & $a_u$          &  0.7434
$\pm$ 0.009 &  0.7772 $\pm$ 0.009 & 0.79167 $\pm$ 0.0106  & 0.79176 $\pm$ 0.0099\\
           & $b_u$         &  3.8907 $\pm$ 0.040 &
 4.0034 $\pm$ 0.033  &  4.02637 $\pm$ 0.0402  & 4.02685 $\pm$ 0.0327    \\
           & $c_u  $                 &  0.1620             &
 0.1000    & 0.0940    &  0.0940         \\
           & $d_u$             & 1.2100             &
1.1400         & 1.1100    &  1.1100     \\
\hline $d_v$      & $a_d$         &  0.7369
$\pm$ 0.040 &  0.7858 $\pm$ 0.043  &  0.80927 $\pm$ 0.0621  & 0.80927 $\pm$ 0.0407  \\
           & $b_d$          &  3.5051 $\pm$ 0.225 &
 3.6336 $\pm$ 0.244   & 3.76847 $\pm$ 0.3499 & 3.76858 $\pm$ 0.2278 \\
           & $c_d  $              & 0.3899             &
0.1838       & 0.1399  & 0.1399      \\
           & $d_d$             & -1.3700             &
-1.2152     & -1.1200   &  -1.1200     \\
\hline \multicolumn{2}{|c|}{$\Lambda_{\rm QCD}^{\rm N_f=4}$, MeV}
&
263.8 $\pm$ 30 & 239.9 $\pm$ 27  &  241.44 $\pm$ 29 &  241.45 $\pm$ 27 \\
\hline \hline \multicolumn{2}{|c|}{$\chi^2 / ndf$} & 523/546 =
0.9578 &
506/546 = 0.9267 & 491.07/546 = 0.8994   &  491.12/546 = 0.8995 \\
\hline \hline
\end{tabular}
\caption{\label{tab:param}{\sf Parameter values of the NLO,
N$^2$LO from Ref.~\cite{Khorramian:2008yh} and N$^3$LO nonsinglet
QCD fit at $Q_0^2 = 4~ \mbox{GeV}^2$ for Pad\'e [1/1] and Pad\'e
[0/2].}}
%\begin{ruledtabular}
%\end{ruledtabular}
\end{table*}
\vspace{1 cm}

In Table (\ref{tab:param}) we summarize the NLO, N$^2$LO, and
N$^3$LO with using Pad\'e [1/1] and [0/2] fit results for the
parameters of the parton densities $xu_v(x,Q^2_0)$,
$xd_v(x,Q^2_0)$ and $\Lambda_{\rm QCD}^{\rm N_f =4}$. The values
without error have been fixed after a first minimization since the
data do not constrain these parameters well enough. In this table
we also  compare the N$^3$LO  results with the NLO and N$^2$LO
results from Ref.\cite{Khorramian:2008yh}. The results show a good
compatibility between Pad\'e [1/1] and [0/2] approximations in
4--loop order. The resulted value of $\chi^2/ndf$ is 0.9578 at
NLO,  0.9267 at N$^2$LO, and 0.8994 and 0.8995 for Pad\'e [1/1]
and [0/2] respectively at N$^3$LO. Our results for the covariance
matrix of the  N$^3$LO nonsinglet QCD fit for Pad\'e [1/1] and
[0/2] are presented in Table(\ref{tab:cov}).

\renewcommand{\arraystretch}{0.9}
\begin{table*}
\begin{tabular}{|c||c|c|c|c|c|}
\hline \hline
{\bf N$^3$LO} Pad\'e[1/1] & $a_{u}$ & $b_{u}$ & $a_{d}$ & $b_{d}$& $\Lambda_{\rm QCD}^{\rm N_f=4}$ \\
\hline
 $a_{u}$ &~{\bf 1.13$\times$10$^{-4}$}&  &  &  &  \\
\hline
 $b_{u}$        &~2.35$\times$10$^{-4}$ &{\bf ~1.62$\times$10$^{-3}$} &  &  &  \\
\hline
 $a_{d}$        &~1.09$\times$10$^{-4}$ &~-1.59$\times$10$^{-3}$ & {\bf ~3.86$\times$10$^{-3}$} &  & \\
\hline
 $b_{d}$        &~1.67$\times$10$^{-4}$ &~-8.84$\times$10$^{-3}$ &~2.11$\times$10$^{-2}$ &~{\bf 1.23$\times$10$^{-1}$} & \\
\hline
 $\Lambda_{QCD}^{(4)}$        &~1.71$\times$10$^{-4}$ &~-3.49$\times$10$^{-4}$ &~5.04$\times$10$^{-4}$ ~&2.61$\times$10$^{-3}$ &~{\bf 8.65$\times$10$^{-4}$} \\
\hline \hline
{\bf N$^3$LO} Pad\'e[0/2]  & $a_{u}$ & $b_{u}$ & $a_{d}$ & $b_{d}$& $\Lambda_{\rm QCD}^{\rm N_f=4}$ \\
\hline
 $a_{u}$ & {\bf ~0.98$\times$10$^{-4}$} &  &  &  &  \\
\hline
 $b_{u}$        &~1.83$\times$10$^{-4}$ &~{\bf 1.07$\times$10$^{-3}$} &  &  &  \\
\hline
 $a_{d}$        &~-5.07$\times$10$^{-5}$ &~-6.01$\times$10$^{-4}$ &~{\bf 1.66$\times$10$^{-3}$} &  & \\
\hline
 $b_{d}$        &~-1.11$\times$10$^{-4}$ &~-3.30$\times$10$^{-3}$ &~8.58$\times$10$^{-3}$ &~{\bf 5.19$\times$10$^{-2}$} & \\
\hline
 $\Lambda_{QCD}^{(4)}$        &~1.59$\times$10$^{-4}$ &~-1.99$\times$10$^{-4}$& 1.94$\times$10$^{-4}$ &~8.07$\times$10$^{-4}$ &~{\bf 7.53$\times$10$^{-4}$} \\
\hline \hline
\end{tabular}

\caption{\label{tab:cov}{\sf Our results for the covariance matrix
of the N$^3$LO nonsinglet QCD fit for Pad\'e [1/1] and [0/2]  at
$Q_0^2 = 4~\GeV^2$ by using MINUIT\cite{MINUIT}.}}
\end{table*}

\begin{figure}
\vspace{2 cm}
\begin{center}
\includegraphics[angle=0, width=9.5cm]{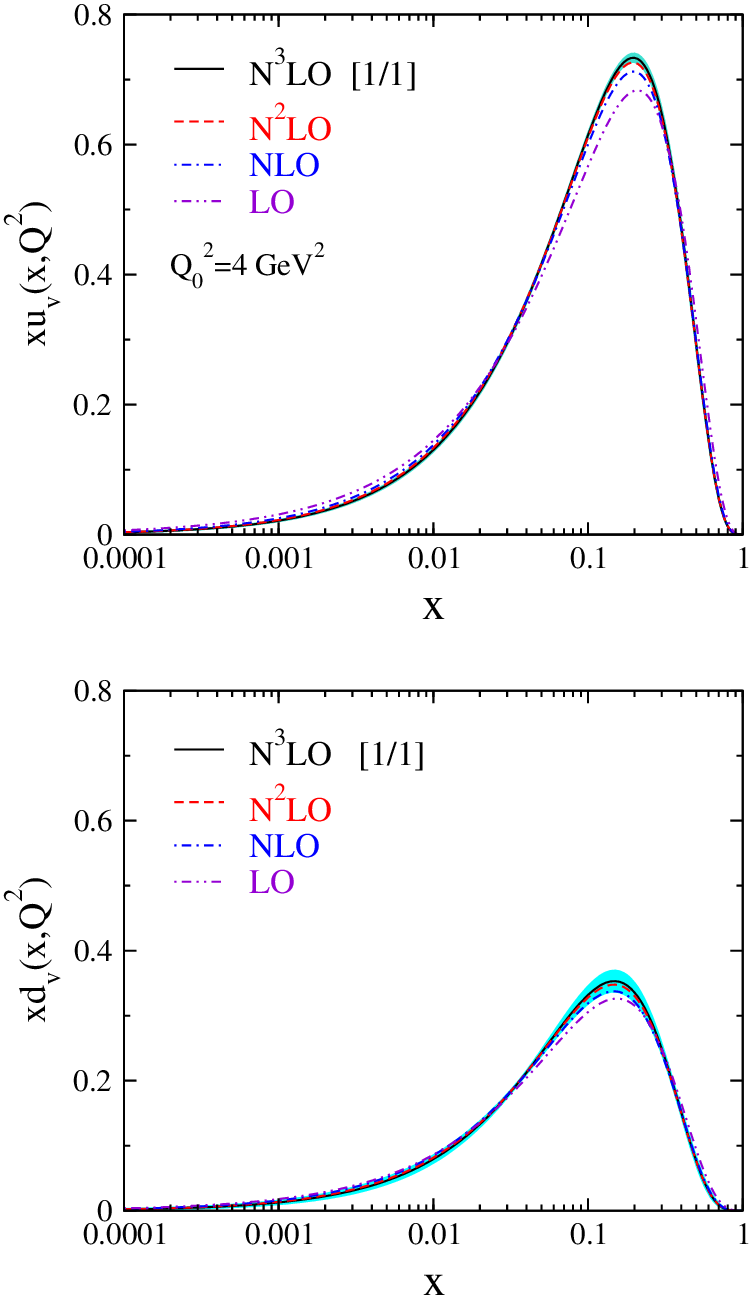}
\end{center}
\vspace{0 cm} {\sf \caption{\label{fig:2} {\sf The parton
densities $xu_v$ and $xd_v$ up to 4-loop (Pad\'e [1/1]) at the
input scale $Q_0^2 = 4.0~{\rm{\GeV^2}}$ (solid line) compared with
results obtained from N$^2$LO analysis (dashed-- line), NLO
analysis (dash--dotted line), and LO analysis(dash--dott--dotted
line)~\cite{Khorramian:2008yh}. The shaded areas represent the
fully correlated one $\sigma$ statistical error bands.}}}
\end{figure}

Figure~(\ref{fig:2}) illustrates our fit results for
$xu_v(x,Q^2_0)$, $xd_v(x,Q^2_0)$ at $Q_0^2 = 4 \GeV^2$ up to
N$^3$LO and for Pad\'e [1/1] with correlated errors. In this
figure our results for N$^3$LO  compared with results obtained
from \cite{Khorramian:2008yh} at LO, NLO, and N$^2$LO QCD
analysis. The shaded areas represent the fully correlated one
$\sigma$ statistical error bands.

\begin{figure}
\vspace{2 cm}
\includegraphics[angle=0, width=16 cm]{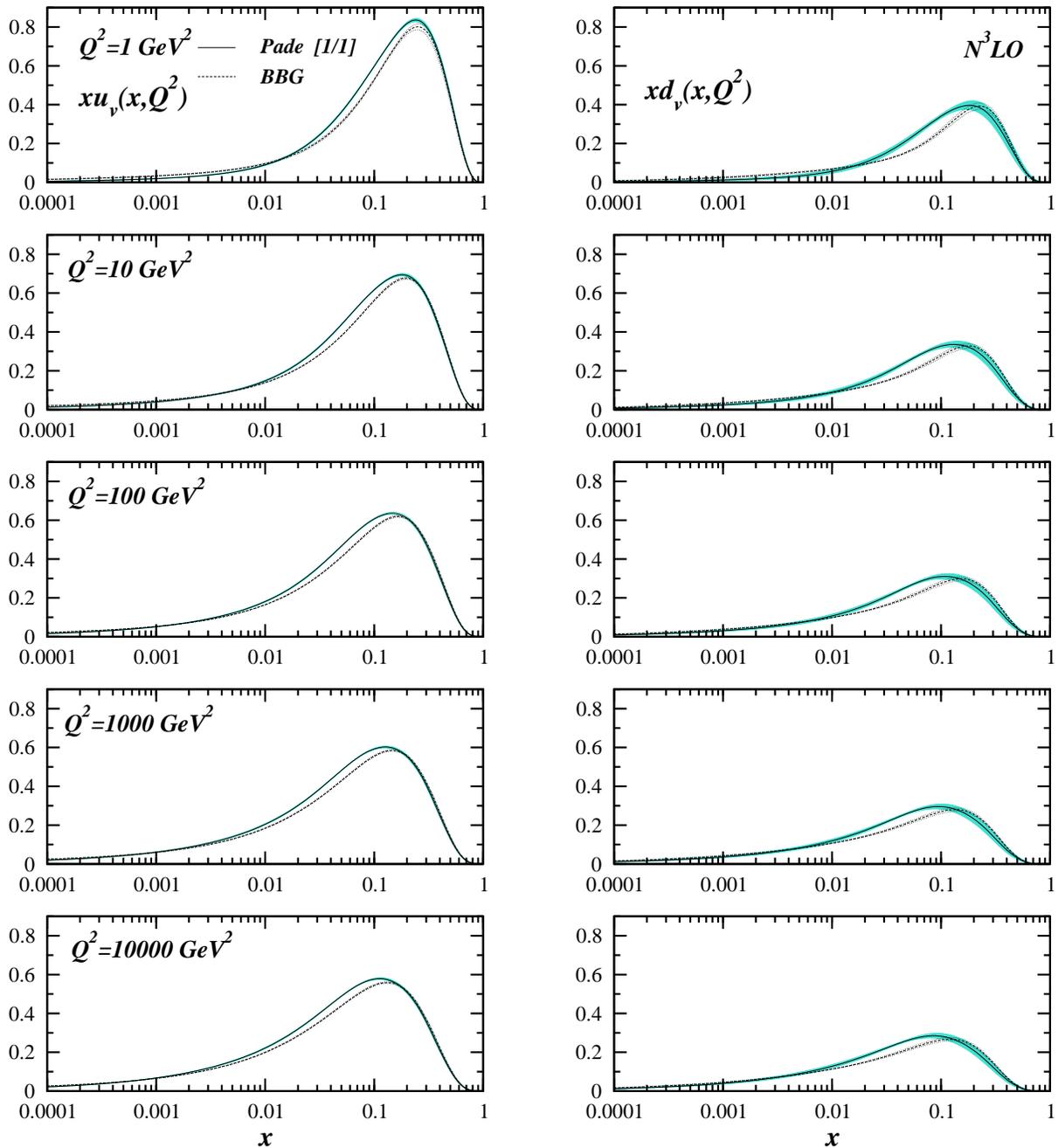}
\begin{center}
\end{center}
{\sf \caption{\label{fig:3} {\sf The parton densities $xu_v$ and
$xd_v$ at N$^3$LO evolved up to $Q^2 = 10000~{ \GeV^2}$ (solid
lines) compared with results obtained by
%A05 (dashed
%line)~\cite{Alekhin:2005gq},
BBG (dashed line)~\cite{Blumlein:2006be}
%, and MRST (dashed--dotted-dottedline)~\cite{Martin:2007bv,MRST04}
.}}}
\end{figure}

In Fig.~(\ref{fig:3}) we show the evolution of the valence quark
distributions $xu_v(x,Q^2)$ and $xd_v(x,Q^2)$ from $Q^2 = 1
\GeV^2$ to $Q^2 = 10^4 \GeV^2$ in the region  $x\in [10^{-4},1]$
at N$^3$LO. In this figure we also compared our results with the
nonsinglet QCD analysis from \cite{Blumlein:2006be}. With rising
values of $Q^2$ the distributions flatten at large values of $x$
and rise at low values.

Another way to test the N$^3$LO fit results is comparison of low
order moments of the distributions $u_v(x,Q^2), d_v(x,Q^2),$ and
$u_v(x,Q^2)- d_v(x,Q^2)$. In Table~\ref{tab:table6} we present the
lowest nontrivial moments of these distributions at $Q^2 = Q_0^2$
in N$^3$LO and compare to the respective moments obtained for the
parameterizations \cite{Blumlein:2006be}.

 We should note that the unknown parameters
 are correlated and almost depend on the method of the QCD fits.
 We believe that the source of the small
difference between the results of our analysis and reported
results in \cite{Blumlein:2006be} is the kind of the different
method of the QCD analysis. We used the Jacobi polynomial method
as an expansion method to do QCD fits but they used the exact
inverse Mellin technique to obtain some unknown parameters. We
also found that the results of Pad\'e [1/1] and [0/2] in 4-loop
level are almost the same.

\renewcommand{\arraystretch}{0.9}
\begin{table*}
\begin{tabular}{|c|c|c|c|c|}
\hline \hline  $f$ & N & BBG~\cite{Blumlein:2006be} & N$^{3}$LO Pad\'e[1/1] & N$^{3}$LO Pad\'e[0/2] \\
\hline
$u_{v}$ & $2$ & $0.3006\pm 0.0031$ & $0.30757\pm 0.0026$ & $0.30806\pm 0.0028$\\
& $3$ & $0.0877\pm 0.0012$ & $0.08771\pm 0.0011$& $0.08781\pm 0.0012$ \\
& $4$ & $0.0335\pm 0.0006$ & $0.03320\pm 0.0006$& $0.03323\pm 0.0006$ \\
\hline
$d_{v}$ & $2$ & $0.1252\pm 0.0027$ & $0.12450\pm 0.0024$& $0.12495\pm 0.0025$ \\
& $3$ & $0.0318\pm 0.0009$ & $0.03040\pm 0.0008$& $0.03012\pm 0.0008$ \\
& $4$ & $0.0106\pm 0.0004$ & $0.00992\pm 0.0004$& $0.00993\pm 0.0005$ \\
\hline
$u_{v}-d_{v}$ & $2$ & $0.1754\pm 0.0041$ & $0.18305\pm 0.0036$& $0.18310\pm 0.0038$ \\
& $3$ & $0.0559\pm 0.0015$ & $0.05767\pm 0.0013$ & $0.05769\pm 0.0014$ \\
& $4$ & $0.0229\pm 0.0007$ & $0.02329\pm 0.0007$& $0.02329\pm
0.0007$
\\ \hline \hline
\end{tabular}
\normalsize
\vspace{2mm} \caption{\label{tab:table6}{\sf Comparison of low
order moments from our nonsinglet N$^3$LO QCD analysis at $Q_0^2 =
4~\mbox{GeV}^2$ with the N$^3$LO analysis from
Ref.~\cite{Blumlein:2006be}. }}
\end{table*}

\begin{figure}
\begin{center}
\includegraphics[angle=0, width=13.0cm]{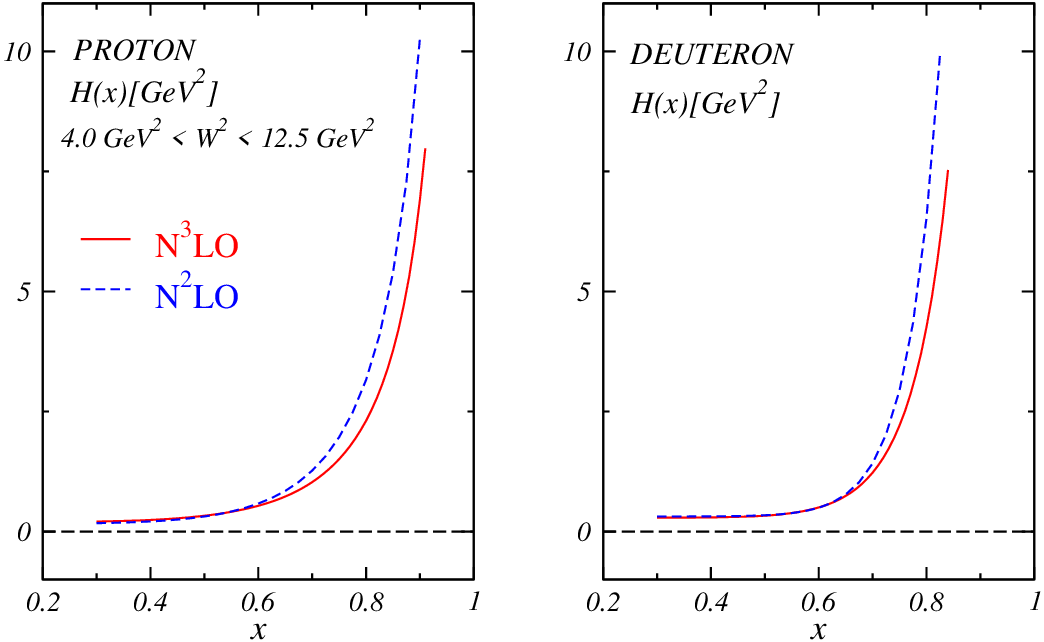}
\end{center}
{\sf \caption{\label{fig:4} The higher twist coefficient $h(x)$
for the proton and deuteron data as a function of $x$ in N$^3$LO
(solid line) compared with results obtained by N$^2$LO (dashed
line) \cite{Khorramian:2008yh}
% and BBG (dashed--dotted line) \cite{Blumlein:2008kz}
. }}
\end{figure}

To perform higher twist QCD analysis of the nonsinglet world data
in N$^3$LO, we consider the $Q^2 \geq 4 \GeV^2, 4<W^2 < 12.5
\GeV^2$ cuts. The number of data points in the above range for
proton and deuteron is 279 and  278, respectively.
 The extracted distributions for $h(x)$ in N$^3$LO are depicted in
Fig.(\ref{fig:4}) for the nonsinglet case considering scattering
off the proton and deuteron target. According to our results the
coefficient $h(x)$ grows towards large $x$.To compare, we also
present the reported results of the early N$^2$LO analysis
\cite{Khorramian:2008yh} in Fig.(\ref{fig:4}). Also in this figure
HT contributions have the tendency to decrease form N$^2$LO to
N$^3$LO. This effect was observed for the first time in the case
of fits of $F_3$ DIS $\nu N$ data in \cite{Kataev:1997nc} and then
studied in more detail in \cite{Kataev:1999bp,Kataev:2001kk}.

This similar effect was also observed in the fits of $F_2$ charge
lepton-nucleon DIS data
\cite{Yang:1999xg,Blumlein:2006be,Gluck:2006yz,Blumlein:2008kz}.
% Note that the results for $h(x)$ in LO are not
%presented in the BBG model \cite{Blumlein:2006be,Blumlein:2008kz}.
In Ref. \cite{Gluck:2006yz}, the functional form for $h(x)$ is
chosen by
\begin{equation}
h(x) = a\left( \frac{x^b}{1-x} -c\right)\,, \label{eq.hx}
\end{equation}
and it is possible to compare $h(x)$ results in N$^2$LO and
N$^3$LO . In Table \ref{tab:hx} we present our results for $a, b,
c$ in the above equation.

\begin{table*}
\begin{tabular}{cccc}
\hline\hline & $a$ & $b$ & $c$ \\ \hline Proton & $~~1.015~~$ &
$~~3.928~~$ & $~~-0.193~~$ \\ \hline Deuteron & $4.481$ & $7.759$
&
$~~-0.064~~$ \\
\hline
\end{tabular}
\vspace{2mm} \caption{\label{tab:hx}{\sf Our results for h(x)
function according to Eq.~(\ref{eq.hx}) and for N$^3$LO.}}
\end{table*}

As seen from Fig.(\ref{fig:4})  $h(x)$ is widely independent of
the target comparing the results for deeply inelastic scattering
off protons and deuterons.

\section{Discussion}
A study \cite{Khorramian:2008yh} of the available world data on
deep-inelastic lepton-proton and lepton-deuteron scattering
provided a determination of the valence quark parton densities and
$\alpha_s$ in wide ranges of the Bjorken scaling variable $x$ and
$Q^2$ up to 3-loop. In the nonsinglet case, where heavy flavor
effects are negligibly small, the analysis can be extended to
4-loop level, i.e. to QCD in N$^3$LO perturbative expansion.

 The analysis was performed
using the Jacobi polynomials method to determine the parameters of
the problem in a fit to the data. A new aspect in comparison with
previous analysis is that we determine the parton densities and
the QCD scale up to N$^3$LO by using  the Jacobi polynomial
expansion method and using Pad\'e approximations. The benefit of
this approach is the possibility to determine nonsinglet parton
distributions analytically and not numerically. In Ref.
\cite{Program:summary} we arrange the MATHEMATICA program to
extract $xu_v(x,Q^2)$ and $xd_v(x,Q^2)$ up to the 4-loops.

In this analysis we adopt the $\bar{d}-\bar{u}$ distribution at
$Q_0^2=4$ GeV$^2$ from
Refs.~\cite{Gluck:2006yz,Blumlein:2006be,ref19,Blumlein:2004pr},
which gives a good description of the Drell-Yan dimuon production
data \cite{E866}. The nonsinglet regime is manifesting itself at
$x\geq 0.1$ as the rule. In this regime, when we changed the sea
distribution from the other groups,  the value of $\chi^2$,
valence distributions, $\Lambda$ and $\alpha_s$ varied, but only
slightly. For example, we used the $\bar{d}-\bar{u}$ distribution
from
\cite{Gluck:2007ck,Martin:2007bv,JimenezDelgado:2008hf,Martin:2009iq}
and we found that the value of  $\chi^2$ varies by about 3\%
 and $\Lambda$ by about 1\%-2\%.

In the QCD analysis we parameterized the strong coupling constant
$\alpha_s$ in terms of four massless flavors determining
$\Lambda_{\rm QCD}$. Up to N$^3$LO results fitting the data, are
\begin{eqnarray}
\Lambda_{\rm QCD}^{(4)} &=& 213.2 \pm 28
\; \mbox{MeV},~~{\tt LO}, \nonumber \\
\Lambda_{\rm QCD}^{(4)} &=& 263.8 \pm 30 \;
\mbox{MeV},~~{\tt NLO}, \nonumber \\
\Lambda_{\rm QCD}^{(4)} &=& 239.9 \pm 27 \;
\mbox{MeV},~~{\tt N^2LO},\nonumber \\
\Lambda_{\rm QCD}^{(4)} &=& 241.4 \pm 29 \; \mbox{MeV},~~{\tt
N^3LO}.
\end{eqnarray}
These results can be expressed in terms of $\alpha_s(M_Z^2)$:
\begin{eqnarray}
\alpha_s(M_Z^2) &=& 0.1281 \pm 0.0028,~~{\tt LO},  \nonumber \\
\alpha_s(M_Z^2) &=& 0.1149 \pm 0.0021,~~{\tt NLO}, \nonumber \\
\alpha_s(M_Z^2) &=& 0.1131 \pm 0.0019,~~{\tt N^2LO}, \nonumber \\
\alpha_s(M_Z^2) &=& 0.1139 \pm 0.0020,~~{\tt N^3LO}.
\end{eqnarray}
Note that in above results we use the matching between $n_f$ and
$n_{f+1}$ flavor couplings calculated in Ref.
\cite{Chetyrkin:1997sg}. We adopt this prescription to be able to
compare our results with other measurement of $\Lambda_{\rm QCD}$.

 The $\alpha_s(M_Z^2)$ values can be
compared with results from other QCD analysis of inclusive
deep--inelastic scattering data in N$^2$LO
\normalsize
\begin{center}
\begin{tabular}{rcllll}
 & & & & & \\
A02   \cite{A02}:      & & $\alpha_s(M_Z^2)$=0.1143 & $\pm$0.0014 &   \\
GRS  \cite{Gluck:2006yz}:&  & $\alpha_s(M_Z^2)$=0.111 &  &
\\
 MRST03 \cite{MRST03}:     & &$\alpha_s(M_Z^2)$= 0.1153 & $\pm$0.0020 &    \\
 SY01(ep)  \cite{SY}:  & & $\alpha_s(M_Z^2)$=0.1166 & $\pm$0.0013 &                \\
 SY01($\nu$N) \cite{SY}:& & $\alpha_s(M_Z^2)$=0.1153 & $\pm$0.0063 &                \\
A06 \cite{Alekhin:2006zm}: & &  $\alpha_s(M_Z^2)$=0.1128 & $\pm$
0.0015 &
\\
BBG \cite{Blumlein:2006be}:  & & $\alpha_s(M_Z^2)$=0.1134  & \footnotesize{$\begin{array}{c} +0.0019 \\
-0.0021 \end{array}$} &\\
BM07   \cite{Brooks:2006wh}:      & & $\alpha_s(M_Z^2)$=0.1189 &
$\pm$0.0019 &
\\
KPS00($\nu N$) \cite{Kataev:1999bp}:      & &
$\alpha_s(M_Z^2)$=0.118 &$\pm$ 0.002 ~($stat$)$\pm$ 0.005~($syst$) \\
&&& $\pm$
0.003~($theory$) &  \\
KPS03($\nu N$) \cite{Kataev:2001kk}:      & &
$\alpha_s(M_Z^2)$=0.119 &$\pm$ 0.002 ~($stat$)$\pm$ 0.005~($syst$)\\
&&&  $\pm$ 0.002 ~($threshold$)
$^{+0.004}_{-0.002}$~ ($scale$) &  \\
KT08 \cite{Khorramian:2008yh}:  & & $\alpha_s(M_Z^2)$=0.1131  & \footnotesize{$\begin{array}{c} \pm 0.0019  \end{array}$} &\\
\end{tabular}
\end{center}
\normalsize The N$^3$LO values of $\alpha_s(M_Z^2)$ can also be
compared with results from other QCD analysis \noindent
\normalsize
\begin{center}
\begin{tabular}{rcllll}
 & & & & & \\
BBG \cite{Blumlein:2006be}:  & & $\alpha_s(M_Z^2)$=0.1134  & \footnotesize{$\begin{array}{c} +0.0019 \\
-0.0021 \end{array}$} &\\
\end{tabular}
\end{center}
\normalsize and with the value of the world average $0.1189 \pm
0.0010$ \cite{Bethke:2006ac} and also the current world average
%-------------------------------------------------------------------------
\begin{eqnarray}
 \alpha_s(M_Z^2) = 0.1184 \pm 0.0007~,
\end{eqnarray}
%-----------------
which has been extracted in \cite{Bethke:2009jm} very recently. It
seems that our results confirm that the value of $\alpha_s(M_Z^2)$
from DIS turns out to be sizably below the world average. In this
case, it would be useful to find out which set of data is mainly
responsible for the low value of $\alpha_s(M_Z^2)$. We will try to
see which subset makes $\alpha_s(M_Z^2)$ particularly small in a
future work.

We hope our results of QCD analysis of structure functions in
terms of Jacobi polynomials
 could be able to describe more complicated hadron structure functions. We also
hope to be able to consider massive quark contributions by using
the structure function expansion in terms of the Jacobi
polynomials.

\newpage
\begin{acknowledgments}
We are especially grateful to G. Altarelli  for guidance and
critical remarks. A.N.K. is grateful to F. Olness and J.
Bl\"umlein for useful discussions and constructive comments. H.K.
is grateful to S.~Moch for his guidance and discussion.  We would
like to thank Z. Karamloo and M. Ghominejad for reading the
manuscript of this paper. A.N.K is grateful to TH-PH division at
CERN for their hospitality while he visited there and  amended
this paper. A.N.K. thanks Semnan University for partial financial
support of this project. We acknowledge the School of Particles
and Accelerators, Institute for Research in Fundamental Sciences
(IPM) for financially supporting this project.
\end{acknowledgments}

\newpage

\newpage

\newpage

\newpage
\newpage

\normalsize

%%%%%%%%%%%%%%%%%%%%%%%%%%%%%%%%%%%%%%%%%%%%%%%%%%%%%%%%%%%%%%%%%%%%%%%
%%%%%%%%%%%%%%%%%%%%%%%%%%%%%%%%%%%%%%%%%%%%%%%%%%%%%%%%%%%%%%%%%%%%%%%
%   Bibliography
%%%%%%%%%%%%%%%%%%%%%%%%%%%%%%%%%%%%%%%%%%%%%%%%%%%%%%%%%%%%%%%%%%%%%%%
\clearpage

\bibliography{apssamp}% Produces the bibliography via BibTeX.

\end{document}